\documentclass[10pt,conference]{IEEEtran}
\IEEEoverridecommandlockouts
\usepackage{cite}
\usepackage{amsmath,amssymb,amsfonts}
\usepackage{graphicx}
\usepackage{tabularx}
\usepackage{multirow}
\usepackage{textcomp}
\usepackage{xcolor}
\usepackage{algorithm}
\usepackage{algpseudocode}
\usepackage{float}
\usepackage{booktabs}
\usepackage{graphicx}
\usepackage{caption}
\usepackage{amsmath}
\usepackage{array}
\usepackage{tabularx}
\usepackage{makecell}
\DeclareUnicodeCharacter{2223}{\ensuremath{\vert}}
\DeclareUnicodeCharacter{2229}{\ensuremath{\cap}}
\DeclareUnicodeCharacter{222A}{\ensuremath{\cup}}

\usepackage{listings}
\usepackage{xcolor}

\definecolor{codeblue}{RGB}{0,0,200}
\definecolor{codegray}{RGB}{110,110,110}
\definecolor{codered}{RGB}{190,50,50}

\lstdefinestyle{cpppattern}{
    language=C++,
    basicstyle=\ttfamily\scriptsize,
    keywordstyle=\color{codeblue}\bfseries,
    commentstyle=\color{codegray},
    stringstyle=\color{codered},
    frame=single,
    rulecolor=\color{black},
    backgroundcolor=\color{white},
    numbers=none,
    showstringspaces=false,
    breaklines=true,
    breakatwhitespace=false,
    columns=fullflexible,
    keepspaces=true,
    tabsize=2,
    xleftmargin=6pt,
    xrightmargin=6pt,
    framexleftmargin=2pt,
    framexrightmargin=2pt,
    aboveskip=4pt,
    belowskip=5pt
}
\def\BibTeX{{\rm B\kern-.05em{\sc i\kern-.025em b}\kern-.08em
    T\kern-.1667em\lower.7ex\hbox{E}\kern-.125emX}}
\makeatletter
\renewcommand\section{\@startsection {section}{1}{\z@}%
    {-2.5ex \@plus -1ex \@minus -.2ex}%
    {1.3ex \@plus.2ex}%
    {\normalfont\normalsize\bfseries\raggedright}}
\makeatother

\begin{document}

\title{\fontsize{24pt}{24pt}\selectfont ShikumiMiner: Mining Recurring Implementation Patterns in AI Codebases\\
%{\footnotesize \textsuperscript{*}Note: Sub-titles are not captured in Xplore and should not be used}
%\thanks{Identify applicable funding agency here. If none, %delete this.}
}

\author{
  \IEEEauthorblockN{Afsana Tasnim}
  \IEEEauthorblockA{
    \textit{University of Texas at Arlington}\\
    Arlington, TX, USA\\
    axt0868@mavs.uta.edu
  }
  \and
  \IEEEauthorblockN{Saif Uddin Mahmud}
  \IEEEauthorblockA{
    \textit{University of Texas at Arlington}\\
    Arlington, TX, USA\\
    sxm8931@mavs.uta.edu
  }
  \and
  \IEEEauthorblockN{Md Erfan}
  \IEEEauthorblockA{
    \textit{University of Alabama}\\
    Tuscaloosa, AL, USA\\
    merfan@crimson.ua.edu
  }
  \and
  \IEEEauthorblockN{Md Rayhanur Rahman}
  \IEEEauthorblockA{
    \textit{University of Alabama}\\
    Tuscaloosa, AL, USA\\
    mrahman87@ua.edu
  }
}
\maketitle

\begin{abstract}
\color{black}
Large language models are paving the way towards
innovation by understanding, analyzing, summarizing and generating content in the modern world. Currently there are thousands of LLM projects developed by engineers in open-source repositories. However, whether these LLM projects have underlying patterns or not remains a question. Exploring these underlying patterns will give new dimensions to the developers who aim to develop these LLM projects. In this paper, we propose ShikumiMiner, a static-analysis framework that combines Abstract Syntax Tree (AST) and Control Flow Graph (CFG) features to detect and compare recurring implementation patterns in C++ local LLM codebases. We analyze ten GitHub open-source repositories and classify functions into seven study-specific categories using a multi-label Random Forest model. Studying these patterns can provide useful insights for developers aiming to design LLM applications. 
\end{abstract}


\section{\textbf{Introduction}}
\color{black}
%\subsection{Main Topic}
Large Language Models (LLMs) are a widespread phenomenon in today's world. Even though these AI-powered innovations have exceptional capabilities like summarization, content generation, complex query answering, code analysis, etc, there are significant concerns regarding the source code used to develop some of the LLM projects. Whether these LLM projects follow any structural or execution pattern or not remains a question that needs to be solved if we want to understand LLM projects better. Code analysis is an option for detecting such patterns in LLM projects. Although prior work has studied ASTs, CFGs, data-flow graphs, and other program representations for code analysis, \cite{ma2026exploring}, \cite{jiang2021treebert}, \cite{guo2022unixcoder}, \cite{lazar2014clone}, \cite{wahler2004clone}, \cite{baxter1998clone},  limited work has examined how these techniques can be used to characterize recurring implementation patterns in C++ local LLM codebases. %This gap motivates our work on ShikumiMiner, which analyzes local LLM implementations by extracting
%function-level syntactic and control-flow evidence from source code.

%\subsection{Others}
Previous studies have shown a systematic empirical analysis of the extent to which large language models can perform fundamental code analysis tasks in a zero-shot setting \cite{ma2026exploring}. While conducting the study, authors have considered three core dimensions of program comprehension namely syntax parsing, static semantic inference, and dynamic reasoning. The authors examine nine representative tasks, including AST generation, expression matching, CFG construction, call graph generation, data dependency analysis, taint analysis, pointer analysis, equivalent mutant detection, and flaky test reasoning in order to evaluate these dimensions. GraphCodeBERT, a pre-trained model for programming languages, considers the inherent structure of code \cite{guo2020graphcodebert}. By using data flow in the pre-training stage, the authors learn code representation from source code and code structure. Their analysis shows that code structure and newly introduced pre-training tasks can improve GraphCodeBERT \cite{guo2020graphcodebert}. 

%\subsection{Gap}
However, the study \cite{ma2026exploring} also identifies several limitations: when tasks require deeper semantic or runtime reasoning, LLM performance degrades substantially. Their outputs depend largely on prompt design and representation format. It is also challenging to evaluate the generated ASTs, CFGs, and semantic graphs because multiple structurally different outputs may still be partially correct. Overall, the paper concludes that LLMs have promising potential for assisting code analysis. However, their reliability limitations require validation mechanisms and hybrid integration with deterministic analyzers. The paper's \cite{guo2020graphcodebert} limitations include its reliance on accurate data-flow extraction, which may be difficult for incomplete, dynamically typed, or complex real-world code; its focus on data flow while excluding richer structures such as control flow or full AST semantics; and evaluation mainly on benchmark datasets, so its robustness in large, noisy, industrial codebases remains less certain.

In this work, we aim to address these limitations by developing an approach named ShikumiMiner. Rather than relying solely on syntactic structure or control-flow information \cite{li2017code}, \cite{nguyen2015graph}, \cite{rabinovich2017abstract}, \cite{white2016deep}, \cite{mou2016convolutional}, our approach seeks to combine both perspectives in order to capture richer program characteristics, including code organization, execution paths, and structural dependencies. 
%By integrating AST- and CFG-based analysis for C++ local LLM systems, this work aims to provide a practical static-analysis framework for characterizing recurring implementation patterns in real-world local LLM
%codebases. 

%\subsection{Our contributions}

In this paper, we make the following contributions:

\begin{itemize}
    \item \textbf{We propose ShikumiMiner, an AST-CFG-based static-analysis framework for extracting function-level structural and control-flow features from C++ local LLM codebases.}

    \item \textbf{We formalize recurring implementation structures in local LLM codebases as detectable pattern categories using combined syntactic and control-flow evidence.}

    \item \textbf{We evaluate ShikumiMiner on ten open-source C++ local LLM repositories and report function-level and project-level pattern distributions.}

    \item \textbf{We analyze project-level similarities and feature-pattern relationships to show how local LLM implementations differ in their structural and execution characteristics.}
\end{itemize}

Prior work has extensively studied ASTs, CFGs, data-flow graphs, and program-dependence graphs for code representation, vulnerability detection, clone detection, defect prediction, and pretrained code-model analysis. However, these studies mainly focus on general software-engineering tasks, neural code representation, or the internal capabilities of pretrained code models. In contrast, ShikumiMiner focuses on real C++ local LLM implementations and uses combined AST and CFG analysis to identify recurring structural and execution patterns. Therefore, ShikumiMiner does not claim to invent AST or CFG analysis; rather, its contribution lies in applying and integrating these representations for domain-specific pattern tracing in local LLM systems.

\section{\textbf{Background and Motivation:}}

\subsection{\textbf{Why Abstract Syntax Tree:}}
An Abstract Syntax Tree (AST) is a hierarchical tree data structure that represents the logical syntactic structure of the source code \cite{wan2018improving}, \cite{alon2018code2seq}. Previous works have analyzed Abstract Syntax Trees to derive meaningful information from source code \cite{hu2018deep}, \cite{jiang2007deckard}, \cite{zhang2019novel}, \cite{alon2019code2vec}, \cite{bui2021treecaps}, \cite{tufano2019empirical}. There are numerous works on using AST for code analysis \cite{allamanis2017learning}, \cite{alon2020structural}, \cite{brockschmidt2018generative}, \cite{buratti2020exploring}, \cite{hellendoorn2019global}, \cite{yin2017syntactic}, \cite{xiong2018identifying}, \cite{zhang2019novel}, \cite{ben2018neural}, \cite{yamaguchi2012generalized}. Bug patterns were also detected by studying the abstract syntax tree of source code \cite{tasnim2018inferring}, \cite{pradel2017deep}, and by using AST-parsed code expressions together with contextualized source-code embeddings \cite{karampatsis2020scelmo}. Combining source-code context and AST structure improves performance compared with using either technique alone \cite{zugner2021language}. Previously, thorough structural analysis has been conducted in order to provide an interpretation of pre-trained language models for source code \cite{kim2021code}. Code analysis was done using CodeBERT \cite{feng2020codebert} and GraphCodeBERT \cite{guo2020graphcodebert} as the source code. Their findings suggest that it may be helpful to incorporate the syntax structure of code into the process of pre-training for better code representation \cite{wan2022they}.

Authors have proposed CodeT5, a unified pre-trained encoder-decoder Transformer model for both code understanding and code generation tasks \cite{wang2021codet5}. 
%The paper argues that many earlier code models are either encoder-only or decoder-only. Encoder-only models are better for understanding but weaker for generation. Decoder-only models are better for generation but weaker for understanding. However, unified model can support both code understanding and generation tasks and allow for multi-task learning, which is why CodeT5 uses the T5 encoder-decoder architecture and introduces identifier-aware pre-training tasks, including identifier tagging and masked identifier prediction. 
Specifically, they convert the PL segment into an Abstract Syntax Tree (AST) and extract the node types for each code token. Their unified model can support both code understanding and generation tasks and allow for multi-task learning. Investigation has been done on what kinds of code knowledge are captured inside pretrained code models such as CodeBERT, GraphCodeBERT, PLBART, CodeT5, and CodeGPT2  \cite{troshin2022probing}. Instead of only evaluating these models on downstream tasks, the paper uses probing tasks to test whether their hidden representations contain information about code syntax, namespaces, data flow, variable naming, variable misuse, semantic equivalence, and readability. 
%The goal is to understand whether these models truly encode meaningful source-code properties
%or only perform well because of surface-level patterns.

\subsection{\textbf{Why Control Flow Graph:}}
A Control Flow Graph (CFG) is a visual or mathematical representation of all possible execution paths through a program. Analyzing the Control Flow Graph (CFG) lets us understand both what the code is built from and how it behaves when it runs. Control Flow Graphs (CFGs) have been used for source-code analysis and code-quality assessment \cite{tufano2018deep}, \cite{zhao2018deepsim}, \cite{xu2017neural}, \cite{jiang2020comparecfg}. CFG-based graph representations have also been used for vulnerability detection, where vulnerability-related structural features are extracted and compared to identify potentially vulnerable code \cite{cui2020vuldetector}. Function similarity was computed based on the CFGs \cite{eschweiler2016discovre}. A combination of control flow and data flow was also used as the basis of similarity metrics to find functional similarity on code \cite{zhao2018deepsim}.  

Many existing deep learning vulnerability detectors rely heavily on token-based transformer models, which can be computationally expensive and can suffer from limited generalization or fail to capture deeper program semantics \cite{chakraborty2021deep}. In order to address this, DeepDFA converts source code into a control flow graph (CFG) and represents each CFG node using an abstract dataflow embedding based on vulnerability-relevant properties such as API calls, data types, constants, and operators \cite{steenhoek2024dataflow}. The paper reports that DeepDFA trains much faster than transformer based models, works well with smaller training datasets, and can improve performance when combined with large language models \cite{steenhoek2024dataflow}. 

\begin{table*}[t]
\centering
\caption{Study-specific implementation-pattern taxonomy used by ShikumiMiner.}
\label{tab:pattern_taxonomy}
\scriptsize
\renewcommand{\arraystretch}{1.15}
\setlength{\tabcolsep}{5pt}

\begin{tabularx}{\textwidth}{
    >{\raggedright\arraybackslash}p{3.0cm}
    >{\raggedright\arraybackslash}p{6.0cm}
    >{\raggedright\arraybackslash}X
}
\hline
\textbf{Pattern} &
\textbf{Definition} &
\textbf{Typical Evidence} \\
\hline

Training Pipeline &
Functions performing training- or optimization-related operations. &
Optimizer steps, backward-propagation logic, gradient-related calls, and loss computation. \\

Interactive Inference &
Functions supporting prompt processing, decoding, and token-generation workflows. &
Generation calls, decoding logic, prompt handling, and iterative token production. \\

Advanced Sampling &
Functions implementing token-selection and sampling strategies. &
Top-$k$, top-$p$, temperature, repeat-penalty, and related sampling operations. \\

Model Loading and Validation &
Functions responsible for loading model data and validating model-related state or inputs. &
Model/file loading calls, validation checks, initialization branches, and failure-handling paths. \\

Memory Management &
Functions performing memory allocation, deallocation, buffer handling, or cache-related memory operations. &
Allocation/deallocation calls, buffer operations, memory initialization, and cache-memory handling. \\

Context Management &
Functions maintaining or updating model, session, decoding, or execution context. &
Context initialization, state updates, cache/context traversal, and repeated state handling. \\

Quantization Workflow &
Functions handling quantized representations or quantization-related operations. &
Quantization calls, quantized data types, conversion logic, and quantized model processing. \\

\hline
\end{tabularx}
\end{table*}

\subsection{\textbf{Analyzing AST and CFG:}}
There have been many works comprising of both AST and CFG \cite{tufano2018deep}, \cite{wang2020detecting}, \cite{zhou2019devign}, \cite{li2021vulnerability}, \cite{yamaguchi2014modeling} for code analysis. Previous studies have revealed that analyzing both AST and CFG can reveal important information about source code \cite{wan2019multi}. GraphCode2Vec, a generic code embedding approach combines lexical or syntactic information with program dependence information. The main problem it addresses is that many existing code embedding methods mainly rely on syntax, such as tokens or AST paths, while ignoring deeper semantic information such as control flow, data flow, and method-call dependencies \cite{ma2022graphcode2vec}. To address this limitation, GraphCode2Vec converts Java programs into the Jimple intermediate representation, extracts lexical information through tokenization and BiLSTM-based instruction embedding, and captures semantic information by constructing program dependence graphs using static analysis with Soot. However, the approach also has some challenges. Since it relies on Soot and Jimple, it is mainly designed for Java and is not suitable for other programming languages. Its effectiveness also depends on the accuracy of the extracted program dependence graphs; if the static analysis misses or misrepresents dependencies, the learned embedding may be affected. 

Tree-Transformer is a neural model for learning program representations directly from syntax trees such as ASTs or CSTs  \cite{wang2023learning}. The paper argues that existing tree-based and graph-based models have several weaknesses: many GNNs capture only local neighborhood information, many tree models propagate information only bottom-up, and several approaches either ignore sibling order or require changing the original tree structure. To address these issues, Tree-Transformer uses a tree-structured multi-head attention mechanism to model both parent-child and sibling relationships. A key limitation of this paper is that this approach preserves AST structure well, but it does not explicitly incorporate CFG-level execution semantics.

The main challenges and limitations of the paper \cite{troshin2022probing} are that probing only shows whether information is extractable from model representations, not whether the model actually uses that
information in real software engineering tasks. The experiments are mainly limited to Java and Python, leaving languages such as C++ less explored. The main challenges and limitations are that the paper \cite{kim2021code} focuses mostly on syntax, not deeper program semantics such as control flow, data flow or semantic equivalence. Also, its analysis is based on correlations between attention, embeddings, and AST structures; this shows that syntax information is present or recoverable, but it does not prove that the model uses this information when solving real software engineering tasks. The main challenges and limitations are that CodeT5 \cite{wang2021codet5} still mainly represents code as a token sequence, even though it uses AST-derived identifier labels during pre-training. This means it does not fully model deeper program structures such as CFGs, data-flow graphs, control dependencies, or runtime behavior. 
%Its performance also depends heavily on the quality of identifier extraction from parsers such as Tree-sitter; parser errors or language-specific parsing limitations can affect the pre-training signal.

%\subsection{\textbf{Why ShikumiMiner:}}
%Previous studies have analyzed the ability of LLMs on code analysis \cite{ma2026exploring}. In ShikumiMiner, we analyze LLM code itself by generating ASTs, constructing CFGs, and applying machine-learning-based pattern detection to identify recurring structural and execution patterns in local LLM implementations. Prior work has applied ASTs, CFGs, and related program representations to general software-engineering tasks, but comparatively little work has used them to characterize recurring implementation concerns in C++ local LLM systems. ShikumiMiner addresses this gap by analyzing the source code of local LLM implementations themselves, using AST-derived syntactic evidence and CFG-derived control-flow evidence to study recurring concerns such as inference, sampling, loading, memory management, context handling, training, and quantization.

\section{\textbf{Proposed Methodology:}}
In this study, an implementation pattern refers to a recurring code-level structure associated with a specific implementation behavior in C++ local LLM systems. ShikumiMiner characterizes these patterns using AST-derived syntactic evidence and CFG-derived control-flow evidence. We use a study-specific taxonomy comprising seven categories: Training Pipeline, Interactive Inference, Advanced Sampling, Model Loading and Validation, Memory Management, Context Management, and Quantization Workflow. These categories are not proposed as new software design patterns; rather, they provide a domain-specific framework for organizing and comparing recurring implementation patterns in local LLM codebases.

\begin{figure}[htbp]
\centerline{\includegraphics[width=0.5\textwidth]{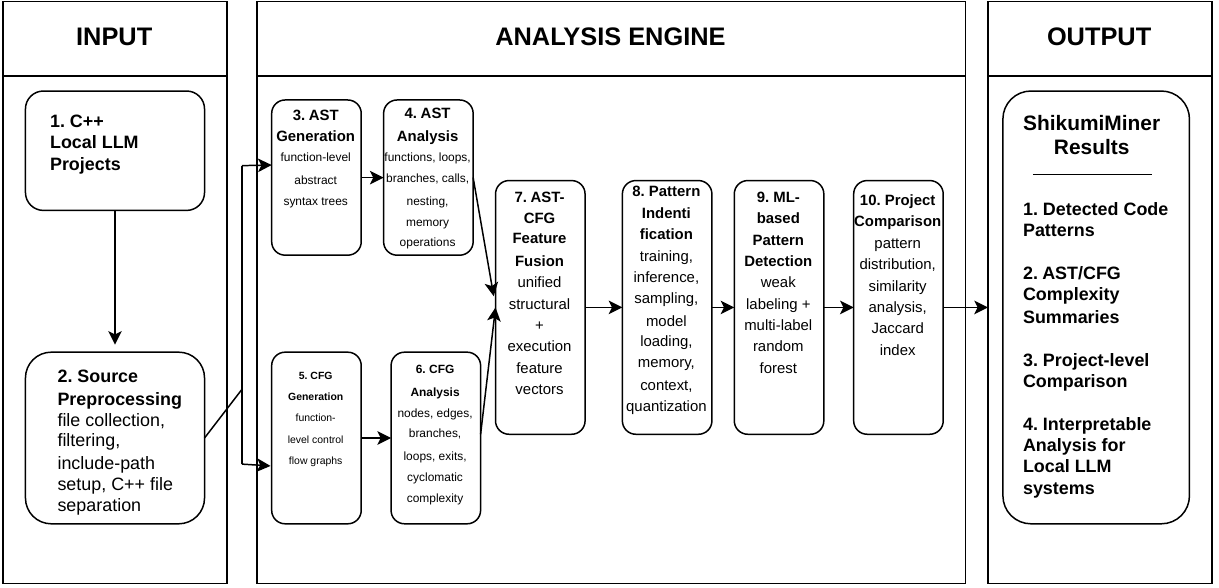}}
\caption{A general framework of ShikumiMiner Approach}
\label{fig1}
\end{figure}

Source preprocessing excludes files unrelated to the core implementation so that the subsequent analysis focuses on code associated with model loading, training, inference, sampling, memory handling, context management, quantization, and deployment-related behavior. Figure \ref{fig1} summarizes the overall ShikumiMiner workflow.

\subsection{\textbf{Abstract Syntax Tree (AST) Generation:}}\label{SCM}

After source preprocessing, ShikumiMiner parses each retained C++ source file using Clang LibTooling to construct an Abstract Syntax Tree. The AST provides a function-level representation of the program’s syntactic structure, including declarations, control statements, function calls, memory-related operations, and domain-specific calls associated with model loading, sampling, training, and quantization. ShikumiMiner traverses these ASTs to extract structural features such as function-call count, loop count, branch count, variable declarations, memory operations, and AST depth. These features provide the syntactic evidence used in subsequent implementation-pattern analysis.

\subsection{\textbf{Control Flow Graph (CFG) Generation:}}\label{SCM}
For each extracted function, ShikumiMiner constructs a Control Flow Graph using Clang's CFG analysis. While the AST represents syntactic structure, the CFG captures the possible execution flow within a function, including branches, loops, back edges, exits, and error-handling paths. ShikumiMiner extracts control-flow features such as the number of CFG nodes, edges, branch nodes, loop nodes, exit nodes, back edges, cyclomatic complexity, loop depth, and validation paths. These features provide execution-oriented evidence for characterizing branching, iteration, and control-flow complexity in subsequent implementation-pattern analysis.

\subsection{\textbf{Feature Fusion and Pattern Detection:}}\label{SCM}

After AST and CFG extraction, ShikumiMiner combines the function-level syntactic and control-flow features described in Sections III-A and III-B into a unified feature vector. Let $f$ denote a function, $A_f$ its AST-based feature vector, and $C_f$ its CFG-based feature vector. The fused representation is defined as

\begin{equation}
X_f = [A_f \parallel C_f]
\end{equation}

where $\parallel$ denotes vector concatenation. The fused representation enables ShikumiMiner to jointly examine syntactic and control-flow characteristics and to evaluate whether these two sources of evidence provide complementary information for implementation-pattern detection.

\subsubsection{\textbf{Weak-Label Generation}}

ShikumiMiner formulates implementation-pattern detection as a multi-label problem because a single function may express more than one implementation pattern. For example, a function may simultaneously contain model-loading and memory-management behavior. For large-scale classifier training, ShikumiMiner generates weak labels using domain-specific rules over the extracted AST and CFG features. These rules correspond to the seven implementation-pattern categories defined in Table~I. For example, optimizer or backward-propagation evidence may indicate Training Pipeline, prompt-processing and token-generation evidence may indicate Interactive Inference, and top-k, top-p, temperature, or repeat-penalty operations may indicate Advanced Sampling. Similarly, model-loading and validation evidence is used for Model Loading and Validation, while quantization-related operations provide evidence for Quantization Workflow. Let $y^{w}_{f,k}$ denote the weak label assigned to function $f$ for pattern category $k$:

\begin{equation}
y^{w}_{f,k} =
\begin{cases}
1, & \text{if the rule for pattern } k \text{ is satisfied},\\
0, & \text{otherwise}.
\end{cases}
\end{equation}

These weak labels provide scalable supervision for the corpus-wide classification experiments. In addition, manually assigned source-level labels are used as independent reference labels for the manual-validation experiment reported in Table~V.

\subsubsection{\textbf{Multi-Label Pattern Classification}}

The fused feature vectors and weak labels are used to train a multi-label Random Forest classifier. ShikumiMiner trains one binary classifier for each implementation-pattern category so that multiple patterns can be assigned to the same function. For a pattern category $k$, the prediction for function $f$ is defined as

\begin{equation}
\hat{y}_{f,k}
=
\mathbb{I}
\left(
\frac{1}{T}
\sum_{t=1}^{T}
h_{t,k}(X_f)
\geq
\tau_k
\right),
\end{equation}

where $T$ is the number of trees, $h_{t,k}(X_f)$ denotes the prediction produced by tree $t$ for pattern $k$, $\tau_k$ is the decision threshold for that pattern, and $\mathbb{I}$ is an indicator function. The complete multi-label prediction for function $f$ is represented as

\begin{equation}
\hat{Y}_f =
[\hat{y}_{f,1}, \hat{y}_{f,2}, \ldots, \hat{y}_{f,K}],
\end{equation}

where $K$ denotes the total number of implementation-pattern categories. Algorithm~2 summarizes weak-label generation and multi-label classifier training.

\subsubsection{\textbf{Project-Level Pattern Profiling}}

After function-level prediction, ShikumiMiner aggregates the detected labels by repository to construct project-level implementation-pattern profiles. These profiles summarize the occurrence of the seven implementation patterns within each analyzed repository and support comparison across local LLM systems. Algorithm~1 summarizes the complete ShikumiMiner pipeline, including source collection, AST and CFG construction, feature extraction, feature fusion, weak-label generation, classifier training, function-level prediction, and project-level aggregation. The resulting profiles are subsequently used to analyze pattern distributions and similarities among the analyzed local LLM repositories. The challenges we try to solve in this paper are:

\begin{algorithm}[t]
\caption{ShikumiMiner Overall Pipeline}
\label{alg:ShikumiMiner}

\textbf{Input:} $\mathcal{R}$: C++ local LLM repositories\\
\textbf{Input:} $\mathcal{P}$: implementation-pattern categories\\
\textbf{Output:} $\hat{Y}$: function-level predictions\\
\textbf{Output:} \textit{Profiles}: project-level pattern profiles
\begin{algorithmic}[1]

%\Statex \textbf{Input:} $\mathcal{R}$: C++ local LLM repositories
%\Statex \textbf{Input:} $\mathcal{P}$: implementation-pattern categories
%\Statex \textbf{Output:} $\hat{Y}$: function-level predictions
%\Statex \textbf{Output:} \textit{Profiles}: project-level pattern profiles

\State $X \gets \langle\,\rangle$
\State $Meta \gets \langle\,\rangle$

\ForAll{$r \in \mathcal{R}$}
    \State $F \gets \textsc{CollectFiles}(r)$
    \State $F \gets \textsc{FilterFiles}(F)$

    \ForAll{$file \in F$}
        \State $AST \gets \textsc{GenAST}(file)$

        \ForAll{$f \in \textsc{Functions}(AST)$}

            \If{$\neg\,\textsc{IsDefinedIn}(f,file)$}
                \State \textbf{continue}
            \EndIf

            \State $cfg \gets \textsc{GenCFG}(f)$

            \If{$cfg=\bot$}
                \State \textbf{continue}
            \EndIf

            \State $A_f \gets \textsc{ASTFeatures}(f)$
            \State $C_f \gets \textsc{CFGFeatures}(cfg)$
            \State $X_f \gets [A_f \parallel C_f]$

            \State $\textsc{Append}(X,X_f)$
            \State $\textsc{Append}(Meta,(r,f))$

        \EndFor
    \EndFor
\EndFor

\State $M \gets \textsc{TrainPatternModel}(X,\mathcal{P})$

\State $\hat{Y} \gets \langle\,\rangle$

\For{$i \gets 1$ \textbf{to} $|X|$}
    \State $\hat{Y}_i \gets \textsc{Predict}(M,X_i)$
    \State $\textsc{Append}(\hat{Y},\hat{Y}_i)$
\EndFor

\State $Profiles \gets \emptyset$

\ForAll{$r \in \mathcal{R}$}
    \State $\hat{Y}_r \gets
    \textsc{ProjectLabels}(\hat{Y},Meta,r)$
    \State $Profiles[r] \gets
    \textsc{Aggregate}(\hat{Y}_r)$
\EndFor

\State \Return $\hat{Y}, Profiles$

\end{algorithmic}
\end{algorithm}

\begin{itemize}
\item \textbf{How can syntactic information extracted from ASTs and behavioral information derived from CFGs be effectively integrated into a unified representation that captures both implementation structure and execution behavior without losing important contextual information?}
\item \textbf{How can graph-based and syntactic features be transformed into meaningful machine-learning representations that effectively capture similarities and differences in implementation and memory-management structures across independently developed LLM systems?}
\item \textbf{How can LLM C++ projects with varying architectures, code sizes, implementation styles, and dependency structures be compared fairly to identify common implementation practices and distinguish project-specific design choices from broadly adopted engineering strategies?}
\end{itemize}

\section{\textbf{Evaluation:}}

In this section, we present the experimental setup and evaluate ShikumiMiner in terms of implementation-pattern detection, feature representation, cross-project generalization, and similarities among local LLM repositories. 

\begin{table*}[t]
\centering
\caption{Summary of the ten analyzed repositories.}
\resizebox{\textwidth}{!}{%
\begin{tabular}{lllrrrrrp{4cm}}
\toprule
\textbf{Repository} &
\textbf{Language} &
\textbf{Source Files} &
\textbf{Parsed Files} &
\textbf{Functions} &
\textbf{CFGs} &
\textbf{Parse Rate} &
\textbf{Main Purpose} \\
\midrule
llama.cpp  & C++ & 330 & 325 & 4846 & 4823 & 98.5\% & Local LLM inference and quantization\\
gemma.cpp & C++ & 38 & 33 & 373 & 373 & 86.8\% & Lightweight Gemma inference engine\\
ONNX Runtime GenAI  & C++ & 145 & 128 & 1717 & 1700 & 88.3\% & ONNX-based generative AI inference\\
OpenVINO GenAI  & C++ & 281 & 252 & 2272 & 2249 & 89.7\% & OpenVINO-based GenAI inference \\
LMDeploy  & C++ & 78 & 77 & 835 & 829 & 98.7\% & LLM deployment and serving\\
gpt2.cpp  & C++ & 1 & 1 & 5 & 4 & 100\% & GPT-2 training or inference logic\\
minchatgpt.cpp  & C++ & 3 & 3 & 37 & 37 & 100\% & Minimal GPT-style inference\\
InferLLM  & C++ & 22 & 22 & 306 & 306 & 100\% & CPU-based local LLM inference\\
SGLang  & C++ & 52 & 51 & 425 & 254 & 98.1\% & High-performance LLM serving\\
DeepSpeed-FastGen & C++ & 70 & 67 & 441 & 380 & 95.7\% & High-throughput text generation\\
\bottomrule
\end{tabular}%
}
%\caption{Summary of the ten analyzed repositories.}
\label{tab:repository_summary}
\end{table*}

\subsection{\textbf{Experimental Setup:}}

We evaluate ShikumiMiner on ten open-source C++ local LLM repositories: \textit{llama.cpp}, \textit{gemma.cpp}, ONNX Runtime GenAI,
OpenVINO GenAI, LMDeploy, \textit{gpt2.cpp},
\textit{minchatgpt.cpp}, InferLLM, SGLang, and
DeepSpeed-FastGen. Documentation, tests, build scripts, and third-party dependencies are excluded. Across the repositories, 11,257 function records are extracted; removing duplicates leaves 11,109 distinct functions, of which 10,807 have valid CFGs. For the ablation study, duplicate fused feature vectors are
collapsed, leaving 7,945 distinct vectors. Table \ref{tab:repository_summary} summarizes the ten analyzed repositories, including their source-file counts, parsed files, extracted functions, valid CFGs, parse rates, and primary purposes.

The multi-label classifier uses a Random Forest with 800 trees, \texttt{min\_samples\_leaf=5}, balanced-subsample class weighting, and random seed 42. One binary classifier is trained per pattern category, and per-pattern thresholds are selected on the training
split using out-of-bag F1. Duplicate vectors are removed to prevent train--test leakage, and weak-label-defining features are withheld from the corresponding classifier.

We evaluate ShikumiMiner using 5-fold stratified cross-validation for AST-only, CFG-only, and fused AST+CFG representations, Leave-One-Project-Out validation for cross-project generalization, and manual validation on 250 functions plus a disjoint sample of 80 functions. Performance is reported using precision, recall,
and F1-score. The implementation uses Clang/LLVM 22.1.8, Python 3.12, and scikit-learn 1.9.0.

\subsection{\textbf{Research Questions:}}

To guide this evaluation, we organize the experiments around the following research questions.

\textbf{RQ1: How does combining syntactic and control-flow evidence affect the detection of recurring implementation patterns in local LLM systems?}

%We answer RQ1 by extracting AST features and CFG features from each C++ function. AST features show what code structures exist, such as loops, calls, branches, memory operations, and sampling logic. CFG features show how the function executes through paths, branches, loops, and exits. Combining both gives stronger evidence for detecting recurring code patterns.

To answer RQ1, we compare AST-only, CFG-only, and fused AST+CFG representations through an ablation study.

\begin{algorithm}[t]
\caption{Weak-Label Generation and Multi-Label Classifier Training}
\label{alg:pattern_detection}

\textbf{Input:} $X$: fused AST-CFG feature vectors\\
\textbf{Input:} $\mathcal{P}$: set of pattern categories\\
\textbf{Output:} $M$: trained multi-label classifier
\begin{algorithmic}[1]

%\Statex \textbf{Input:} $X$: fused AST-CFG feature vectors
%\Statex \textbf{Input:} $\mathcal{P}$: set of pattern categories
%\Statex \textbf{Output:} $M$: trained multi-label classifier

\State $Y^{w} \gets \emptyset$

\ForAll{$X_f \in X$}
    \State $Y_f^{w} \gets [0,\ldots,0]$

    \ForAll{$p_k \in \mathcal{P}$}
        \If{$\textsc{Rule}_k(X_f)=true$}
            \State $Y_{f,k}^{w} \gets 1$
        \EndIf
    \EndFor

    \State $Y^{w} \gets Y^{w} \cup \{Y_f^{w}\}$
\EndFor

\State $M \gets \textsc{TrainClassifier}(X,Y^{w})$

\State \Return $M$

\end{algorithmic}
\end{algorithm}

\begin{table*}[t]
\centering
\caption{Patterns identified across the ten analyzed projects.}
\resizebox{\textwidth}{!}{%
\begin{tabular}{lcccccccc}
\toprule
\textbf{Project} &
\textbf{Training} &
\textbf{Inference} &
\textbf{Sampling} &
\textbf{Loading} &
\textbf{Memory} &
\textbf{Context} &
\textbf{Quantization} &
\textbf{Dominant Pattern} \\
\midrule
llama.cpp  & 43	& 1,267 & 318 &	923 & 394 &	371 & 126 & Interactive Inference \\
gemma.cpp  & 4 & 61	& 11 & 29 & 19 &  31 & 3 & Interactive Inference \\
ONNX Runtime GenAI   & 	0 & 315 & 149 & 98 & 176 & 254 & 2 & Interactive Inference \\
OpenVINO GenAI  & 0 & 463 & 62 & 102 & 93 & 176 & 5 &	Interactive Inference \\
LMDeploy  & 2 & 85 & 10 & 42 & 63 & 63 & 5	& Interactive Inference \\
gpt2.cpp  & 0 & 1 & 0 & 0 & 0 & 0 & 0 & Interactive Inference \\
minchatgpt.cpp   & 0 & 17 & 2 & 1 & 1 & 0 & 1 &	Interactive Inference \\
InferLLM  & 0 & 18 & 7 & 47 & 23 & 2 & 10 & Model Loading and Validation \\
SGLang  & 0 & 19 & 11 & 18 & 12 & 19 & 18 & Interactive Inference \\
DeepSpeed-FastGen & 31 & 25 & 6 & 11 & 36 & 11 & 24 & Memory Management
 \\
\bottomrule
\end{tabular}%
}
%\caption{Patterns identified across the ten analyzed projects.}
\label{tab:optimization_patterns}
\end{table*}

\color{black}
\textbf{RQ2: What recurring implementation patterns appear in C++ local LLM systems?}

We answer RQ2 by analyzing the function-level pattern labels produced by ShikumiMiner and examining which of the seven implementation-pattern categories recur across the analyzed repositories. 
%This indicates that the analyzed C++ local LLM repositories contain recurring structural and execution patterns that can be detected using fused AST-CFG evidence.

\textbf{RQ3: How are detected code patterns distributed across different local LLM repositories?}

We answer RQ3 by aggregating function-level pattern labels at the repository level.
%such as Training Pipeline, Interactive Inference, Advanced Sampling, Memory Management, Context Management, and Quantization Workflow. 

\textbf{RQ4: To what extent does ShikumiMiner generalize across previously unseen local LLM repositories?}

To answer RQ4, we use Leave-One-Project-Out (LOPO) evaluation. 
%This analysis examines the extent to which the AST- and CFG-derived representation captures implementation-pattern evidence that is shared across repositories rather than specific to individual projects. 
%The results provide a measure of cross-project robustness and help characterize how consistently recurring implementation concerns are realized across different local LLM systems.

\textbf{RQ5: How similar are local LLM projects based on their detected implementation-pattern profiles?}

We answer RQ5 by converting each repository into a project-level implementation-pattern profile and comparing these profiles using Jaccard similarity. 
%A higher Jaccard value indicates that two projects share more detected structural and execution patterns, while a lower value indicates that the repositories share fewer detected implementation-pattern categories. This helps identify which local LLM projects follow similar or different design patterns.

\section{\textbf{Results}}\label{SCM}
The experimental results show that ShikumiMiner identifies distinct implementation-pattern profiles across the ten analyzed C++ local LLM repositories, but the distribution of those patterns varies substantially by project. As shown in Table~\ref{tab:optimization_patterns}, Interactive Inference is the dominant category for most repositories, including llama.cpp, gemma.cpp, ONNX Runtime GenAI, OpenVINO GenAI, LMDeploy, gpt2.cpp, minchatgpt.cpp, and SGLang. In contrast, InferLLM is dominated by Model Loading and Validation, while DeepSpeed-FastGen shows Memory Management as its most frequent category. llama.cpp exhibits the broadest overall profile, with substantial instances of inference, model loading, sampling, memory management, context management, quantization, and a smaller amount of training-related code. ONNX Runtime GenAI and OpenVINO GenAI also contain substantial context-management and sampling activity in addition to inference, whereas smaller repositories such as gpt2.cpp and minchatgpt.cpp contain much narrower pattern profiles. 
%Overall, these results indicate that local LLM systems share several recurring implementation patterns, particularly inference, loading, memory handling, and context management, while the relative emphasis of those patterns differs according to repository scope and implementation purpose.

%The ablation results further show that syntactic information contributes more strongly to pattern detection than control-flow information under the current feature representation. The AST-only configuration achieves a macro F1-score of 0.40 and a micro F1-score of 0.46, while the CFG-only configuration performs substantially worse, with macro and micro F1-scores of 0.22 and 0.24, respectively. The fused AST+CFG representation reaches a macro F1-score of 0.39 and a micro F1-score of 0.45, which is slightly lower than AST-only performance. At the per-pattern level, the fused representation performs best for Interactive Inference (F1 = 0.57) and Model Loading and Validation (F1 = 0.55), followed by Memory Management (0.40), Quantization Workflow (0.39), Advanced Sampling (0.34), Context Management (0.27), and Training Pipeline (0.18). These findings suggest that AST-derived syntactic evidence provides most of the discriminative signal for the current taxonomy, while simply concatenating CFG features does not consistently improve classification performance. Control-flow information may still provide useful descriptive evidence for interpreting implementation behavior, but the present results do not show an aggregate classification advantage from naive AST-CFG fusion.

\color{black}

\subsection{\textbf{Representative Examples of Identified Implementation Patterns}}

%New

\noindent\textbf{Training Pipeline:} llama.cpp, ops.cpp:4059
%\hfill{\small (llama.cpp, ops.cpp:4059)}

\noindent\textit{Code that computes gradients, losses or optimizer updates.}

\noindent\textbf{Evidence:} The \texttt{\_back} suffix indicates a
backward (gradient) kernel.

\begin{lstlisting}[style=cpppattern]
void ggml_compute_forward_rms_norm_back(
    const ggml_compute_params * params,
    ggml_tensor * dst) {

    const ggml_tensor * src0 = dst->src[0];

    switch (src0->type) {
        case GGML_TYPE_F32:
            ggml_compute_forward_rms_norm_back_f32(
                params, dst);
            break;
        default:
            GGML_ABORT("fatal error");
    }
}
\end{lstlisting}

%New

%New

\noindent\textbf{Model Loading and Validation: } llama.cpp, openelm.cpp:3 
%\hfill{\small (llama.cpp, openelm.cpp:3)}

\noindent\textit{Code that reads model files, metadata or weights into memory.}

\noindent\textbf{Evidence:} Reads a key from the model file and maps
the layer count to a model-size type.

\begin{lstlisting}[style=cpppattern]
void llama_model_openelm::load_arch_hparams(
    llama_model_loader & ml) {

    ml.get_key(
        LLM_KV_ATTENTION_LAYERNORM_RMS_EPS,
        hparams.f_norm_rms_eps);

    switch (hparams.n_layer()) {
        case 16:
            type = LLM_TYPE_270M;
            break;
        // ... three more sizes
        default:
            type = LLM_TYPE_UNKNOWN;
    }
}
\end{lstlisting}

%New

%New

\noindent\textbf{Interactive Inference: } llama.cpp, llama.cpp:506
%\hfill{\small (llama.cpp, llama.cpp:506)}

\noindent\textit{Code that drives prompt processing, decoding or token generation.}

\noindent\textbf{Evidence:} Detects the chat template and applies it
to the messages.

\begin{lstlisting}[style=cpppattern]
int32_t llama_chat_apply_template(
    const char * tmpl,
    const struct llama_chat_message * chat,
    size_t n_msg, ...) {

    llm_chat_template detected_tmpl =
        llm_chat_detect_template(curr_tmpl);

    int32_t res =
        llm_chat_apply_template(
            detected_tmpl,
            chat_vec,
            formatted_chat,
            add_ass);

    return res;
}
\end{lstlisting}

%New

%New

\noindent\textbf{Memory Management: } InferLLM, device.cpp:34
%\hfill{\small (InferLLM, device.cpp:34)}

\noindent\textit{Code that allocates, frees or copies raw buffers.}

\noindent\textbf{Evidence:} Reuses a freed block when one fits and
otherwise calls \texttt{aligned\_alloc}.

\begin{lstlisting}[style=cpppattern]
void * CPUDevice::allocate(size_t len) {

    auto it = m_free_memory.lower_bound(len);

    if (it != m_free_memory.end() &&
        it->second.size() > 0) {

        void * ptr = it->second.back();
        it->second.pop_back();
        return ptr;
    }

    return aligned_alloc(len);
}
\end{lstlisting}

%New

%New

\noindent\textbf{Advanced Sampling: } 
\hfill{\small (ONNX Runtime GenAI, search.cpp:435)}

\noindent\textit{Code that reshapes token scores before a token is chosen.}

\noindent\textbf{Evidence:} Applies a repetition penalty to the
logits of each beam.

\begin{lstlisting}[style=cpppattern]
void Search_Cpu::ApplyRepetitionPenalty(
    float penalty) {

    if (penalty == 1.0f)
        return;

    for (int i = 0;
         i < params_->BatchBeamSize();
         ++i) {

        ApplyRepetitionPenaltyToLogits(
            beam_token_scores,
            sequence,
            penalty,
            repetition_penalty_visited_);
    }
}
\end{lstlisting}

%New

%New

\noindent\textbf{Context Management: } ONNX Runtime GenAI, decoder\_only.cpp:127
%\hfill{\small (ONNX Runtime GenAI, decoder\_only.cpp:127)}

\noindent\textit{Code that maintains the KV cache, positions or session state.}

\noindent\textbf{Evidence:} Rewinds position inputs and the KV cache
to the accepted sequence length.

\begin{lstlisting}[style=cpppattern]
void DecoderOnly_State::CropToAccepted(
    size_t new_length,
    size_t recurrent_position) {

    position_inputs_->RewindTo(new_length);

    if (kv_cache_)
        kv_cache_->RewindTo(new_length);

    if (recurrent_state_)
        recurrent_state_->CropToPosition(
            recurrent_position);
}
\end{lstlisting}

%New

%New
\noindent\textbf{Quantization Workflow: } DeepSpeed-FastGen, pt\_binding.cpp:139
%\hfill{\small (DeepSpeed-FastGen, pt\_binding.cpp:139)}

\noindent\textit{Code that converts between quantized, full-precision values.}

\noindent\textbf{Evidence:} Dequantizes 4-bit integers to half
precision using per-group scale and minimum values.

\begin{lstlisting}[style=cpppattern]
at::Tensor dequantize_int4_to_half_experimental(
    at::Tensor & data_in,
    at::Tensor & scale_buffer,
    at::Tensor & min_val_buffer,
    ...) {

    launch_dequantize_int4_to_half_experimental(
        (uint8_t *) data_in.data_ptr(),
        (half *) output.data_ptr(),
        ...);

    return output;
}
\end{lstlisting}

%New

\subsection{\textbf{Manual Validation}}
Because the precision, recall, and F1-scores in Table~\ref{tab:full_ablation} measure agreement with rule-based weak labels rather than independent semantic correctness, we performed a manual validation using source-level reference labels. We uniformly sampled 250 functions from the 10,807 distinct functions with valid CFGs across all ten repositories and manually inspected approximately 40 lines of source code for each function while withholding its feature vector, weak label, and ShikumiMiner prediction. Labels were assigned under the multi-label taxonomy in Section III only when a function performed substantive work corresponding to an implementation pattern. Against these independent labels, ShikumiMiner achieved a macro F1 of 0.56 and a micro F1 of 0.53, with 74 true positives, 69 false positives, and 60 false negatives; complete agreement across all seven categories occurred for 145 of 250 functions (58\%), and the 95\% bootstrap confidence interval for micro F1 was [0.46, 0.60]. A second disjoint sample of 80 functions produced a macro F1 of 0.39 and micro F1 of 0.45, while pooling both samples (\(n=330\)) yielded a micro F1 of 0.51 with a 95\% confidence interval of [0.46, 0.57]. These results indicate that performance against weak-supervision targets and performance against independently assigned semantic labels can differ substantially. 
%The validation is limited by the use of a single annotator and the small number of positive examples for rare categories.

\begin{table*}[t]
\centering
\caption{Per-pattern ablation results for AST-only, CFG-only, and fused AST+CFG features.}
\label{tab:full_ablation}

\begin{tabular}{lccccccccc}
\hline
\textbf{Pattern} &
\multicolumn{3}{c}{\textbf{AST Only}} &
\multicolumn{3}{c}{\textbf{CFG Only}} &
\multicolumn{3}{c}{\textbf{AST+CFG}} \\
\cline{2-4} \cline{5-7} \cline{8-10}
&
\textbf{Precision} & \textbf{Recall} & \textbf{F1 Score} &
\textbf{Precision} & \textbf{Recall} & \textbf{F1 Score} &
\textbf{Precision} & \textbf{Recall} & \textbf{F1 Score} \\
\hline

Training Pipeline
& 0.14 & 0.36 &	0.20
& 0.01 & 0.21 &	0.02
& 0.16 & 0.19 &	0.18 \\

Interactive Inference
& 0.56 & 0.59 &	0.58	
& 0.36 & 0.55 & 0.43	
& 0.55 & 0.60 &	0.57 \\

Advanced Sampling
& 0.35 & 0.38 & 0.36	
& 0.08 & 0.25 &	0.12	
& 0.32 & 0.36 & 0.34 \\

Model Loading \& Validation
& 0.69 & 0.46 &	0.55	
& 0.29 & 0.33 & 0.31	
& 0.68 & 0.46 & 0.55 \\

Memory Management
& 0.41 & 0.47 & 0.44	
& 0.17 & 0.30 & 0.22	
& 0.36 & 0.45 & 0.40 \\

Context Management
& 0.20 & 0.44 & 0.27	
& 0.10 & 0.95 & 0.18	
& 0.20 & 0.40 & 0.27 \\

Quantization Workflow
& 0.43	& 0.36 & 0.39	
& 0.56	& 0.18 & 0.28	
& 0.43	& 0.36 & 0.39 \\

\hline
\textbf{Macro Average}
& 0.40 & 0.44 & 0.40	
& 0.22 & 0.40 & 0.22	
& 0.39 & 0.41 & 0.39 \\

\textbf{Micro Average}
& 0.42 & 0.50 & 0.46	
& 0.16 & 0.49 & 0.24	
& 0.42 & 0.49 & 0.45 \\

\hline
\end{tabular}

\end{table*}

\begin{table*}[t]
\centering
\caption{ShikumiMiner evaluated against independent source-derived reference labels ($n=250$).}
\label{tab:manual_validation}
\small
\setlength{\tabcolsep}{6pt}
\renewcommand{\arraystretch}{1.15}
\begin{tabular}{lcccccccc}
\hline
\textbf{Pattern} &
\textbf{Ref. Positives} &
\textbf{Predicted} &
\textbf{TP} &
\textbf{FP} &
\textbf{FN} &
\textbf{Precision} &
\textbf{Recall} &
\textbf{F1} \\
\hline

Training Pipeline
& 4 & 2 & 2 & 0 & 2 & 1.00 & 0.50 &	0.67 \\

Interactive Inference
& 35 & 45 & 22 & 23 & 13 & 0.49 & 0.63 & 0.55 \\

Advanced Sampling
& 10 & 12 &	6 &	6 &	4 &	0.50 & 0.60 & 0.55 \\

Model Loading and Validation
& 22 & 23 &	10 & 13 & 12 & 0.44 & 0.46 & 0.44 \\

Memory Management
& 25 & 24 & 14 & 10 & 11 & 0.58 & 0.56 & 0.57 \\

Context Management
& 31 & 32 &	16 & 16 & 15 & 0.50 & 0.52 & 0.51 \\

Quantization Workflow
& 7 & 5 & 4	& 1 & 3 & 0.80 & 0.57 &	0.67 \\

\hline
\textbf{Macro Average}
& -- & -- & -- & -- & -- &
\textbf{0.62} &
\textbf{0.55} &
\textbf{0.56} \\

\textbf{Micro Average}
& -- & -- &
\textbf{74} &
\textbf{69} &
\textbf{60} &
\textbf{0.52} &
\textbf{0.55} &
\textbf{0.53} \\

\hline
\end{tabular}
\end{table*}

\subsection{\textbf{Ablation Study: AST-Only, CFG-Only, and AST-CFG Fusion}}

Table~\ref{tab:full_ablation} compares three feature configurations: AST-only, CFG-only, and the fused AST+CFG representation. The results show that AST-derived features provide the strongest overall signal for implementation-pattern detection. The AST-only configuration achieves a macro F1-score of 0.40 and a micro F1-score of 0.46, outperforming both CFG-only and the fused representation. In contrast, CFG-only obtains a macro F1-score of 0.22 and a micro F1-score of 0.24, indicating that control-flow characteristics alone are insufficient to distinguish the seven implementation-pattern categories reliably. The fused AST+CFG configuration achieves a macro F1-score of 0.39 and a micro F1-score of 0.45, which is nearly identical to, but slightly lower than, AST-only performance.

At the individual-pattern level, AST-only and AST+CFG perform similarly for several categories. Interactive Inference reaches F1 scores of 0.58 and 0.57, respectively, while Model Loading and Validation remains at 0.55 in both settings. Adding CFG features slightly reduces performance for Advanced Sampling and Memory Management. CFG-only performs substantially worse for most categories; for example, Training Pipeline, Advanced Sampling, and Memory Management achieve F1 scores of 0.02, 0.12, and 0.22, respectively. Although CFG-only attains high recall for Context Management, its low precision results in an F1 score of 0.18. Overall, these results suggest that broad control-flow properties provide limited discriminative power without syntactic context.

Overall, the ablation results show that the syntactic representation carries most of the discriminative information used by ShikumiMiner. Contrary to the expectation that simple feature concatenation would consistently improve detection, the fused AST+CFG representation does not outperform AST-only features at the aggregate level. This finding indicates that the additional CFG dimensions do not necessarily always provide sufficient complementary signal to improve overall performance under the current feature-concatenation strategy. 

\subsection{\textbf{F1 Score:}}
Precision measures the proportion of predicted implementation-pattern instances that are correctly identified, whereas recall measures the proportion of reference instances that are successfully detected by
ShikumiMiner. Because the task is formulated as a multi-label classification problem and the pattern categories exhibit different frequencies, we use the F1-score to provide a balanced assessment of precision and recall. The F1-score is defined as

\begin{equation}
F1 =
2 \times
\frac{\text{Precision} \times \text{Recall}}
{\text{Precision} + \text{Recall}},
\end{equation}

where

\begin{equation}
\text{Precision} =
\frac{TP}{TP + FP},
\end{equation}

and

\begin{equation}
\text{Recall} =
\frac{TP}{TP + FN}.
\end{equation}

\begin{figure}[htbp]
\centerline{\includegraphics[width=0.5\textwidth]{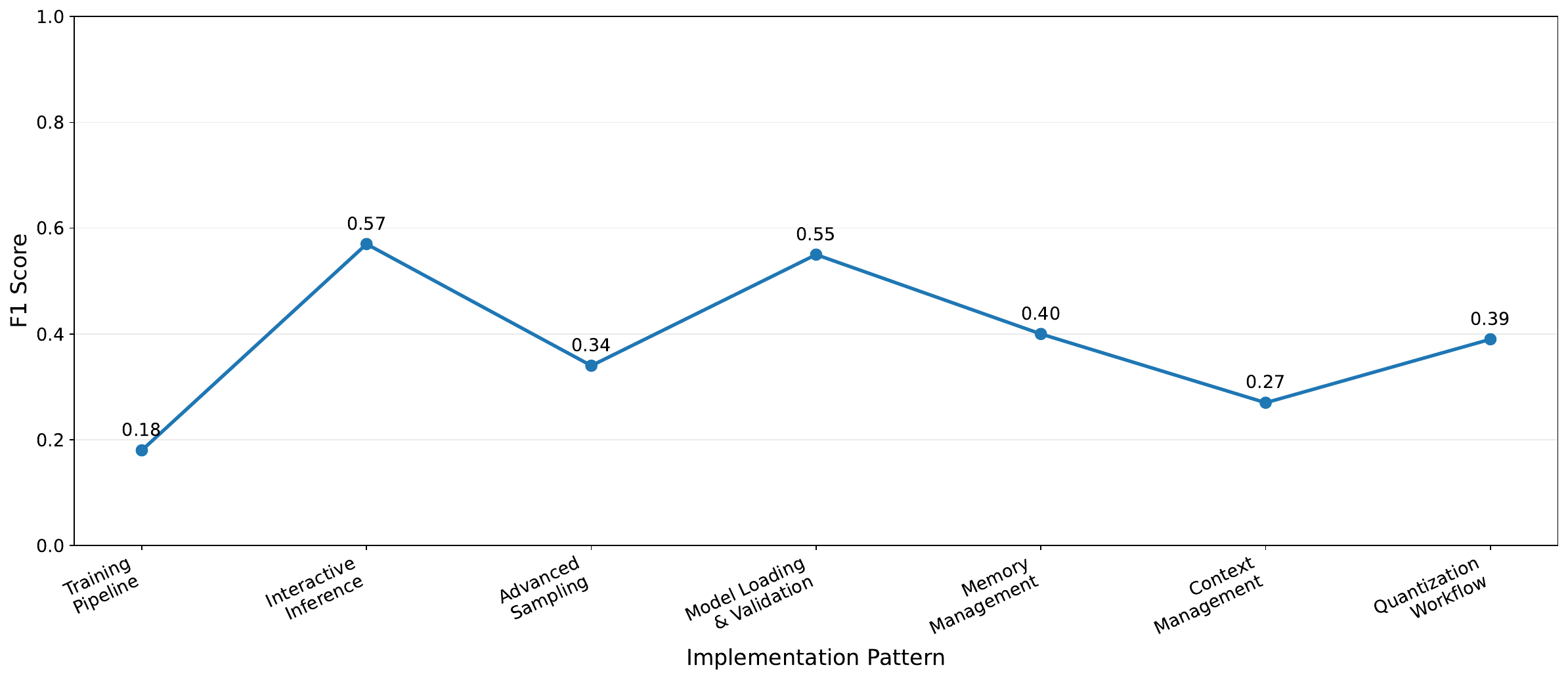}}
\caption{F1 Score Comparison for ShikumiMiner}
\label{fig2}
\end{figure}

%Figure \ref{fig3} presents the per-pattern F1-scores obtained using the fused AST+CFG representation. Interactive Inference achieves the highest F1-score of 0.57, followed closely by Model Loading and Validation at 0.55. Memory Management and Quantization Workflow obtain
%F1-scores of 0.40 and 0.39, respectively, while Advanced Sampling achieves 0.34. Context Management produces a lower F1-score of 0.27, and Training Pipeline is the most difficult category to identify, with an F1-score of 0.18. These results are consistent with the ablation analysis in Table~\ref{tab:full_ablation}, which reports an overall macro F1-score of 0.39 and micro F1-score of 0.45 for the fused representation.

Figure \ref{fig2} summarizes the per-pattern F1-scores for the fused AST+CFG representation. The results differ across patterns, meaning that the AST and CFG features are more useful for identifying some patterns than others. Interactive Inference and Model Loading and Validation are comparatively easier to distinguish because their implementations contain more consistent syntactic evidence, including characteristic function calls and operations. In contrast, Training Pipeline, Context Management, and Advanced Sampling exhibit greater structural overlap with other implementation patterns or contain fewer representative instances, making them more difficult to separate using the current feature representation. In particular, the low F1-score for Context Management may be due to overlapping features with other function types. When similar structural and control-flow characteristics appear across multiple categories, the model can misclassify functions, leading to more false positives and false negatives.

%Overall, the F1-score analysis shows moderate agreement with the weak-supervision targets, with substantial variation across implementation concerns. The results are consistent with the ablation finding that straightforward AST-CFG feature concatenation does not consistently improve performance over AST-only representations. The effectiveness of the fused representation therefore appears to be pattern dependent, suggesting that more selective feature integration or learned fusion strategies may be required to improve discrimination among structurally overlapping categories.

\subsection{\textbf{Leave-One-Project-Out Generalization:}}

ShikumiMiner shows limited cross-project generalization under Leave-One-Project-Out evaluation. Table \ref{tab:lopo_generalization} presents the Leave-One-Project-Out results. Across the ten held-out repositories, the average Macro F1 is 0.112 and the average Micro F1 is 0.167. Performance varies by project, with \texttt{minchatgpt.cpp} obtaining the highest Micro F1 of 0.318 and \texttt{gpt2.cpp} obtaining 0.000, although the latter contains only three test functions. Most repositories achieve Micro F1 values between 0.10 and 0.19. These results suggest that the implementation patterns recur across projects, but their AST- and CFG-level realizations remain strongly repository-specific, limiting transfer to unseen codebases.

%Table \ref{tab:lopo_generalization} evaluates ShikumiMiner using Leave-One-Project-Out (LOPO) validation, where each repository is held out in turn and the model is trained on the remaining nine repositories. Across the ten held-out projects, ShikumiMiner achieves an average Macro F1 of 0.112 and an average Micro F1 of 0.167, indicating limited cross-project generalization. Performance varies across repositories, with minchatgpt.cpp obtaining the highest Micro F1 of 0.318, followed by llama.cpp at 0.274, whereas gpt2.cpp obtains both macro and micro F1-scores of 0.000; however, this result provides only limited evidence, as the held-out set contains just three functions. Most other repositories obtain Micro F1 values between approximately 0.10 and 0.19. These results suggest that although recurring implementation-pattern categories are observed across multiple local LLM systems, the concrete AST- and CFG-level characteristics associated with those patterns differ substantially between repositories. Consequently, models trained on some repositories may not perform well when they are applied to a completely new repository. The results therefore highlight the strong repository-specific nature of implementation-pattern occurrence. The results also suggest that ShikumiMiner may work better on new repositories if it learns from a wider variety of projects and uses features that are not too dependent on one project’s specific code structure or coding style.

\subsection{\textbf{Jaccard Index:}}

To compare project-level implementation-pattern coverage, we represent each repository as the set of pattern categories detected at least once. For repositories $P_i$ and $P_j$, Jaccard similarity is defined as

\begin{equation}
J(P_i,P_j) =
\frac{
\left|\mathrm{Patterns}(P_i)\cap\mathrm{Patterns}(P_j)\right|
}{
\left|\mathrm{Patterns}(P_i)\cup\mathrm{Patterns}(P_j)\right|
}.
\end{equation}

\begin{figure}[htbp]
\centerline{\includegraphics[width=0.5\textwidth]{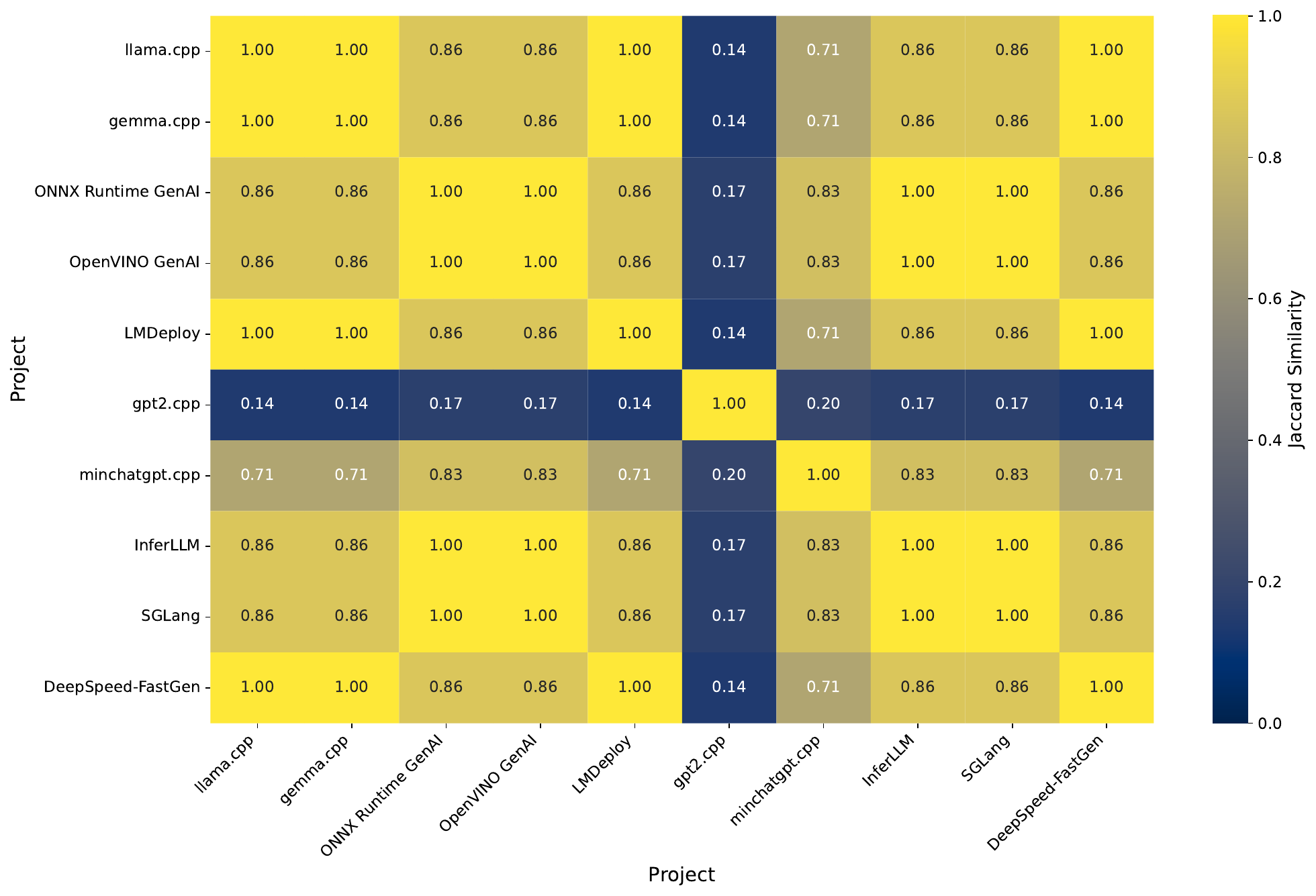}}
\caption{Jaccard Similarity}
\label{fig3}
\end{figure}

A value of $1.0$ indicates identical pattern-category coverage, whereas $0.0$ indicates no shared categories. Figure \ref{fig3} shows the pairwise Jaccard similarities among the repositories.
The results show substantial overlap among the analyzed repositories. \texttt{llama.cpp}, \texttt{gemma.cpp}, \texttt{LMDeploy}, and
\texttt{DeepSpeed-FastGen} contain all seven categories, while ONNX Runtime GenAI, OpenVINO GenAI, InferLLM, and SGLang share the same six categories, with Training Pipeline absent. Their cross-group similarity is approximately $0.86$. \texttt{minchatgpt.cpp} exhibits a narrower five-category profile, whereas \texttt{gpt2.cpp}, containing only Interactive Inference, has the lowest similarity to the other projects.

\begin{table}[t]
\centering
\caption{Leave-One-Project-Out (LOPO) generalization performance of ShikumiMiner.}
\label{tab:lopo_generalization}

\resizebox{\columnwidth}{!}{%
\begin{tabular}{lcccc}
\hline
\textbf{Held-out Repository} &
\textbf{Test Functions} &
\textbf{Patterns} &
\textbf{Macro F1} &
\textbf{Micro F1} \\
\hline
llama.cpp                 & 3,577 & 7 & 0.172 & 0.274 \\
gemma.cpp                 & 335 & 7 & 0.103 & 0.169 \\
ONNX Runtime GenAI        & 1,290 & 6 & 0.134 & 0.189 \\
OpenVINO GenAI            & 1,699 & 6 & 0.115 & 0.166 \\
LMDeploy                  & 647 & 7 & 0.119 & 0.191 \\
gpt2.cpp                  & 3 & 1 & 0.000 & 0.000 \\
minchatgpt.cpp            & 30 & 5 & 0.202 & 0.318 \\
InferLLM                  & 149 & 6 & 0.129 & 0.161 \\
SGLang                    & 217	& 6 & 0.072 & 0.102 \\
DeepSpeed-FastGen         & 283	& 7 & 0.076 & 0.102 \\
\hline
\textbf{Average}          & --    & -- & \textbf{0.112} & \textbf{0.167} \\
\hline
\end{tabular}%
}

\end{table}

%Overall, the Jaccard analysis indicates that most of the analyzed C++ local LLM systems share a common set of recurring implementation patterns despite substantial differences in repository size and implementation purpose. The primary differences arise from whether projects contain specialized functionality such as training or whether they implement only a narrower subset of inference-related patterns. However, because Jaccard similarity considers only the presence or absence of pattern categories, it does not capture differences in their relative frequencies. Two repositories may therefore obtain a Jaccard similarity of $1.00$ even when their pattern distributions differ substantially. Accordingly, Jaccard should be interpreted as a measure of pattern-category coverage similarity rather than similarity in the detailed distribution or architecture of the repositories.

\subsection{\textbf{Spearman Correlation:}}

\begin{table*}[t]
\centering
\caption{Representative case studies illustrating AST and CFG evidence for detected implementation patterns.}
\label{tab:case_studies}
\scriptsize
\renewcommand{\arraystretch}{1.15}
\setlength{\tabcolsep}{3pt}

\begin{tabularx}{\textwidth}{
    >{\raggedright\arraybackslash}p{4.4cm}
    >{\raggedright\arraybackslash}p{1.9cm}
    >{\raggedright\arraybackslash}p{1.9cm}
    >{\raggedright\arraybackslash}p{2.0cm}
    >{\raggedright\arraybackslash}X
}
\hline
\textbf{Project / Function} &
\textbf{Detected Pattern} &
\textbf{AST Evidence} &
\textbf{CFG Evidence} &
\textbf{Interpretation} \\
\hline

\textbf{llama.cpp} \newline
\texttt{llama\_init\_from\_model} \newline
(\texttt{llama-context.cpp:3557})
&
Interactive Inference \newline
(also Memory, Quantization)
&
32 calls, 17 branches, 2 loops, 6 local variables, AST depth 14
&
53 nodes, 83 edges, 31 branch nodes, 2 loop nodes, 2 back edges, cyclomatic complexity 32
&
Context-initialization entry point for inference. The 31 branch nodes show configuration-dependent setup, not straight-line initialization. Its three labels are correct: the function also inspects KV-cache quantization types and allocates the context.
\\
\hline

\textbf{OpenVINO GenAI} \newline
\texttt{StatefulSpeculative} \newline
\texttt{LLMPipeline::generate\_tokens} \newline
(\texttt{fast\_draft\_strategy.cpp:416})
&
Context Management \newline
(also Inference)
&
123 calls, 15 branches, 4 loops, 40 local variables, AST depth 18
&
51 nodes, 71 edges, 21 branch nodes, 4 loop nodes, 4 back edges, cyclomatic complexity 22, loop depth 2
&
Speculative-decoding state handling. The 40 local variables and four loops at depth 2 are the Context Management signature: substantial state is traversed repeatedly, with back edges confirming that it is revisited
rather than initialized once.
\\
\hline

\textbf{llama.cpp} \newline
\texttt{ggml\_opt\_build} \newline
(\texttt{ggml-opt.cpp:322})
&
Training Pipeline \newline
(also Sampling, Memory)
&
93 calls, 7 training-related calls, 30 branches, 4 loops,
29 local variables, AST depth 16
&
101 nodes, 149 edges, 46 branch nodes, 4 loop nodes, 4 back edges, cyclomatic complexity 50, loop depth 1
&
Optimizer graph construction, the densest training function in the corpus. Only 80 of 10,807 functions carry training evidence, and its control flow is indistinguishable from ordinary graph building---which is why the category scores 0.18 in Table~\ref{tab:full_ablation}.
\\
\hline

\end{tabularx}
\end{table*}

We use Spearman rank correlation to examine the relationship between function-level structural characteristics and the amount of detected
implementation-pattern evidence. Each function is treated as one observation, with the number of assigned pattern categories as the response variable. The analyzed metrics include AST depth, function-call
count, loop and branch counts, CFG node and edge counts, CFG branch-node count, CFG loop depth, cyclomatic complexity, and validation-path count.
Spearman correlation is used because these metrics may be skewed and need not follow a normal distribution. The coefficient is defined as

\begin{equation}
\rho_s =
\frac{
\sum_{i=1}^{n}(R_i-\bar{R})(S_i-\bar{S})
}{
\sqrt{\sum_{i=1}^{n}(R_i-\bar{R})^2}
\sqrt{\sum_{i=1}^{n}(S_i-\bar{S})^2}
},
\end{equation}

where $R_i$ and $S_i$ denote the corresponding ranks for function $i$, and $\bar{R}$ and $\bar{S}$ denote their mean ranks.

\begin{figure}[t]
    \centering
    \includegraphics[width=\columnwidth]{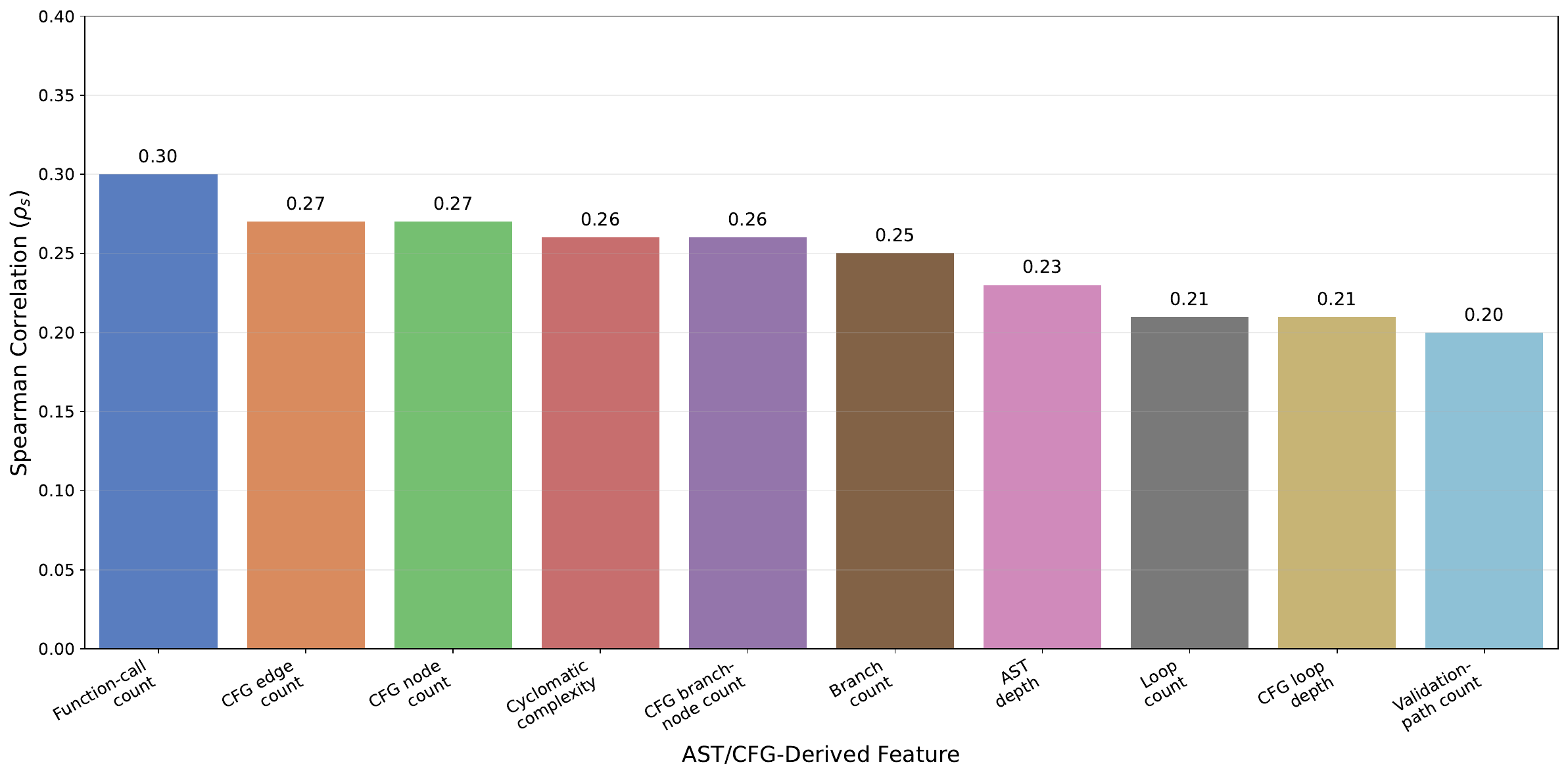}
    \caption{Spearman correlation between AST/CFG-derived features and detected pattern evidence.}
    \label{fig:spearman-correlation}
\end{figure}

Figure~\ref{fig:spearman-correlation} shows weak positive associations across all ten features, ranging from 0.20 to 0.30. Function-call count shows the strongest relationship ($\rho_s=0.30$), followed by CFG edge and node counts ($\rho_s=0.27$). Overall, no individual structural or control-flow metric exhibits a strong monotonic association with detected pattern evidence, indicating that structural size or complexity alone is not a strong proxy for implementation-pattern occurrence.

%The narrow spread across these ten measures is itself informative. No individual structural or control-flow metric shows a strong monotonic association with the number of detected implementation patterns. Larger and more complex functions receive slightly more pattern assignments, but the relationship remains weak, indicating that structural size alone is not a strong proxy for implementation-pattern evidence. 
%This is consistent with the ablation results in Table~\ref{tab:full_ablation}, where the CFG-only configuration achieves substantially lower performance than AST-only and fused representations.

%These correlations should, however, be interpreted as descriptive rather than as independent validation of ShikumiMiner. Several of the analyzed AST- and CFG-derived metrics also contribute directly or indirectly to the feature representation used for pattern detection. Consequently, the observed associations characterize the relationship between ShikumiMiner's input representation and its detected pattern profiles, but they do not establish causality or demonstrate classifier effectiveness. Detection performance is evaluated separately through the ablation, manual-validation, and cross-project experiments.

%\section{\textbf{Case Studies:}}

\section{\textbf{Discussion:}}\label{SCM}
The results show that the seven implementation patterns recur across the analyzed local LLM repositories, although their implementations vary across projects. Syntactic and API-level conventions appear to be more useful than control-flow structure for distinguishing these patterns. Common patterns include inference, model loading, memory management, and context management, while training and quantization appear more often in projects that specifically support those tasks. The ablation results also show that AST features provide stronger discrimination than CFG features, and simple AST+CFG fusion does not improve overall performance. Manual validation and LOPO results further show that ShikumiMiner can struggle when pattern evidence is weak or when implementation styles differ across repositories. Even so, these patterns can help developers building new local LLM systems by serving as implementation references rather than fixed templates. Developers can compare how similar responsibilities are handled in existing systems and select only the patterns that are relevant to their own goals, which can reduce repeated design effort and help avoid overlooking important implementation responsibilities. Table \ref{tab:case_studies} presents representative case studies showing how AST- and CFG-derived features provide evidence for selected implementation-pattern detections across the analyzed repositories.

\section{Threats to Validity}

ShikumiMiner is subject to several limitations. First, the seven-category taxonomy may not capture all implementation strategies used across local LLM repositories. Second, rule-based weak labeling may introduce noise when pattern evidence is ambiguous, incomplete, or distributed across multiple functions. Third, the evaluation is limited to ten open-source C++ repositories, which may restrict generalizability to other LLM systems, industrial projects, or implementations in other languages. 
%Fourth, manual validation was performed by a single annotator on 330 functions in total (250 plus a disjoint sample of 80); larger independently annotated datasets with inter-rater agreement may yield different results. 
Repository size imbalance may also affect aggregate pattern counts and performance, and missing CFGs, particularly for CUDA extensions or specialized build dependencies, may reduce repository-level coverage.

%Our conducted experiments for ShikumiMiner have several threats to validity. First, different projects may organize different patterns in different ways, which is why our proposed pattern taxonomy may not capture all implementation strategies used in local LLM repositories. Second, the use of rule-based weak labeling may introduce noise when pattern evidence is incomplete, ambiguous, or spread across multiple functions. Third, the evaluation is limited to ten open-source C++ local LLM repositories, so the findings may not generalize to all LLM systems, industrial projects, or implementations written in other languages such as Python and Java. Fourth, the reference set comprises 250 functions, together with a disjoint held-out sample of 80 (330 in total), and was annotated by a single annotator, a larger independently annotated dataset with inter-rater agreement might have yielded a different result. In addition, the analyzed repositories vary substantially in size and in the frequency of individual pattern categories. This may give more detected pattern numbers to larger projects in the repository datasets and may influence the overall performance results. In addition, some functions could not be analyzed because valid CFGs were unavailable, particularly in repositories containing CUDA extensions or specialized build dependencies, which may affect repository-level coverage.

\color{black}
\section{\textbf{Conclusion}}

In this study, we presented a static-analysis framework named ShikumiMiner, which characterizes recurring implementation patterns in C++ local LLM repositories. ShikumiMiner successfully identifies these recurring patterns across ten open-source systems through the use of AST- and CFG-derived evidence, while also revealing substantial variation in how these patterns are present across the repositories. The evaluation shows that syntactic evidence provides the strongest discriminative signal, whereas control-flow information contributes primarily to execution-oriented characterization. Overall, ShikumiMiner provides a basis for more interpretable and reproducible comparison of local LLM implementations. As future work, ShikumiMiner paves the way towards richer semantic and cross-project representations of local LLM systems in order to understand the system better and for designing better LLM systems.

%\section{\textbf{Data Availability:}}
%We make our artifact publicly available at https://anonymous.4open.science/r/KTP-C52D/ to support reproducibility and provide access to the materials used in our study.

%\section{\textbf{AI Disclosure:}}
%Claude (Anthropic) was used to assist with code completion during the development. All generated content was reviewed, verified, and revised by the authors. The experimental design, analysis, interpretation of results, and scientific conclusions are entirely the authors’ own work.

%\subsection{\textbf{Acknowledgement}}
%The authors would like to thank Adhvan Inc. for its support throughout this work. This work was financially supported by Adhvan Inc., Tokyo, Japan. 
%The authors thank Adhvan Inc. for its support. A preprint of this work is publicly available on arXiv, and the corresponding replication package and experimental artifacts are available on GitHub.

%\textbf{Data Availability:} All the experimental dataset and code used in this paper are available at https://github.com/AfsanaTasnim/ShikumiMiner for download.

\vspace{12pt}

\bibliographystyle{IEEEtran}
\bibliography{reference_latest}
\end{document}